\documentclass[12pt]{article}
\usepackage{graphicx}
\usepackage{amsfonts}
\usepackage{color}

\def\dalemb#1#2{{\vbox{\hrule height .#2pt
        \hbox{\vrule width.#2pt height#1pt \kern#1pt
                \vrule width.#2pt}
        \hrule height.#2pt}}}

\let\a=\alpha \let\b=\beta \let\g=\gamma \let\d=\delta \let\e=\epsilon
\let\z=\zeta  \let\th=\theta  
\let\l=\lambda \let\m=\mu \let\n=\nu \let\x=\xi \let\p=\pi 
\let\s=\sigma   \let\f=\phi  
 
\let\w=\omega      \let\G=\Gamma  \let\Th=\Theta 
\let\X=\Xi  \let\S=\Sigma  \let\Y=\Psi
 
\let\la=\label  
  
\def\nn{\nonumber} \def\bd{\begin{document}} \def\ed{\end{document}}
\def\ds{\documentstyle} \let\fr=\frac \let\bl=\bigl \let\br=\bigr
\let\Br=\Bigr \let\Bl=\Bigl
\let\bm=\bibitem
\let\na=\nabla
\def\tU{{\widetilde U}}
\let\pa=\partial \let\ov=\overline
\def\ie{{\it i.e.\ }}
\newcommand{\be}{\begin{equation}}
\newcommand{\ee}{\end{equation}}
\def\ba{\begin{array}}
\def\ea{\end{array}}
\def\ft#1#2{{\textstyle{{\scriptstyle #1}\over {\scriptstyle #2}}}}
\def\fft#1#2{{#1 \over #2}}
\def\F#1#2{{ F_{#1}^{(#2)} }}
\def\cF#1#2{{ {\cal F}_{#1}^{(#2)} }}

\def\R{{\bf R}}
\def\sst#1{{\scriptscriptstyle #1}}
\def\oneone{\rlap 1\mkern4mu{\rm l}}
\def\e7{E_{7(+7)}}
\def\td{\tilde}
\def\wtd{\widetilde}
\def\im{{\rm i}}
\def\bog{Bogomol'nyi\ }
\newcommand{\ho}[1]{$\, ^{#1}$}
\newcommand{\hoch}[1]{$\, ^{#1}$}
\newcommand{\bea}{\begin{eqnarray}}
\newcommand{\eea}{\end{eqnarray}}
\newcommand{\ra}{\rightarrow}
\newcommand{\lra}{\longrightarrow}
\newcommand{\Lra}{\Leftrightarrow}
\newcommand{\ap}{\alpha^\prime}
\newcommand{\bp}{\tilde \beta^\prime}
\newcommand{\cB}{{\cal B}}
\newcommand{\cO}{{\cal O}}
\newcommand{\vecx}{\vec{x}}
\newcommand{\vecy}{\vec{y}}
\newcommand{\vecp}{\vec{p}}
\newcommand{\vecq}{\vec{q}}
\newcommand{\tr}{{\rm tr} }
\newcommand{\Tr}{{\rm Tr} }
\newcommand{\NP}{Nucl. Phys. }

\newcommand{\cL}{{\cal L}}
\newcommand{\cA}{{\cal A}}
\newcommand{\cD}{{\cal D}}

\def\sst#1{{\scriptscriptstyle #1}}
\def\0{{\sst{(0)}}}
\def\1{{\sst{(1)}}}
\def\2{{\sst{(2)}}}
\def\3{{\sst{(3)}}}
\def\4{{\sst{(4)}}}
\def\5{{\sst{(5)}}}
\def\6{{\sst{(6)}}}
\def\7{{\sst{(7)}}}
\def\8{{\sst{(8)}}}
\def\ve{\varepsilon}
\def\vf{\varphi}
\def\F{\Phi}
\def\wg{\wedge}

\newcommand{\tamphys}{\it 
}

\newcommand{\auth}{AUTHORS}

\def\thb{\bar{\theta}}
\def\Thb{\bar{\Theta}}
\def\barp{\bar{p}}
\def\barq{\bar{q}}
\def\barc{\bar{c}}
\def\bard{\bar{d}}
\def\e{\epsilon}

\def \bi{\bibitem}
\def \la {\label}

\def \l {\lambda}
\def\foot{\footnote}
\def \tl  {{\tilde \l}}
\def \sql {{\sqrt \l}}
\def \adss {$AdS_5 \times S^5$\ }
\newcommand{\rf}[1]{(\ref{#1})}
\def \ov {\over}

\def\th{\theta}
\def\Th{\Theta}
\def\vth{\vartheta}
\def\btheta{{\bar\theta}}
\def\ttheta{{{\tilde\theta}}}
\def\bttheta{{{\bar\ttheta}}}
\def\vth{\vartheta}

\def\ra{\rightarrow}
\def\N{{\cal N}}
\def\F{{\cal F}}
\def\uM{\underline{M}}
\def\uN{\underline{N}}
\def\uP{\underline{P}}
\def\muu{\underline{\mu}}
\def\nuu{\underline{\nu}}
\def\rou{\underline{\r}}
\def\siu{\underline{\s}}
\def\cc{\circ}
\def\eqv{\equiv}

\def\ni{\noindent}

\def\Ep{E^{{}^{(+)}}}
\def\Em{E^{{}^{(-)}}}

\def\Mp{M^{{}^{(+)}}}
\def\Mm{M^{{}^{(-)}}}

\def \ha{{1\ov 2}}

\def\r{\rho}

\def\Y{{\rm Y}}
\def\X{{\rm X}}
\def\tY{\tilde{\rm Y}}
\def\tX{\tilde{\rm X}}
\def\dY{\dot{\rm Y}}
\def\dX{\dot{\rm X}}

\def \J {\mathcal{J}}
\def \del {\partial}

\def\dF{\dot{F}}
\def\dG{\dot{G}}
\def\df{\dot{f}}
\def \E {{\cal E}}
\def \S {{\cal S}}
\def \J {{\cal J}}

\def\ms{\mathcal{S}}
\def\mj{\mathcal{J}}
\def\soj{\fr{\ms}{\mj}}
\def \R {{\bf R}}
\def \om {\omega}
\def \bE {\bar E}
\def \x {{\cal X}}

\def \bi{\bibitem}
\def \la {\label}

\def \l {\lambda}
\def\foot{\footnote}
\def \tl  {{\tilde \l}}
\def \sql {{\sqrt \l}}
\def \adss {$AdS_5 \times S^5$\ }
\def \ov {\over}

\def \varpi {{\rm w}}

\def\thb{\bar{\theta}}
\def\Thb{\bar{\Theta}}
\def\zb{\bar{z}}
\def\psib{\bar{\psi}}
\def\barp{\bar{p}}
\def\barq{\bar{q}}
\def\barc{\bar{c}}
\def\bard{\bar{d}}
\def\e{\epsilon}
\def\sb{\bar{s}}
\def\wb{\bar{w}}
\def\lb{\bar{\l}}
\def\Jb{\bar{J}}
\def\Nb{\bar{N}}
\def\pab{\bar{\pa}}

\def\At{\tilde{A}}
\def\Bt{\tilde{B}}
\def\Ct{\tilde{C}}
\def\Dt{\tilde{D}}
\def\Et{\tilde{E}}
\def\Ft{\tilde{F}}
\def\Gt{\tilde{G}}
\def\Mt{\tilde{M}}
\def\at{\tilde{a}}
\def\bt{\tilde{b}}
\def\ct{\tilde{c}}
\def\dt{\tilde{d}}
\def\et{\tilde{e}}
\def\ept{\tilde{\e}}
\def\ft{\tilde{f}}
\def\gt{\tilde{g}}
\def\ola{\overleftarrow}
\def\ora{\overrightarrow}
\def\alt{\tilde{\a}}

\def\dh{\hat{d}}
\def\bh{\hat{b}}
\def\deh{\hat{\d}}
\def\mh{\hat{m}}
\def\nh{\hat{n}}

\def\ah{\hat{a}}
\def\eh{\hat{e}}
\def\Eh{\hat{E}}
\def\eph{\hat{\e}}
\def\ph{\hat{p}}
\def\Ah{\hat{A}}
\def\alh{\hat{\a}}
\def\beh{\hat{\b}}
\def\gah{\hat{\g}}
\def\muh{\hat{\m}}
\def\roh{\hat{\r}}
\def\sih{\hat{\s}}
\def\nuh{\hat{\n}}
\def\thh{\hat{\th}}
\def\dh{\hat{d}}
\def\ih{\hat{i}}
\def\jh{\hat{j}}
\def\kh{\hat{k}}
\def\ph{\hat{p}}
\def\qh{\hat{q}}
\def\rh{\hat{r}}
\def\sh{\hat{s}}
\def\deh{\hat{\d}}
\def\uh{\hat{u}}
\def\vh{\hat{v}}
\def\wh{\hat{w}}
\def\lah{\hat{\l}}
\def\Ch{\hat{C}}
\def\Omh{\hat{\Omega}}
\def\rh{\hat{r}}
\def\sh{\hat{s}}
\def\that{\hat{t}}
\def\lah{\hat{\l}}

\def\dgg{\dagger}

\def\ps{\rlap{\, /}\;\,p }
\def\ks{\rlap{\, /}\;\,k }
\def\pas{\rlap{\, /}\;\,\pa }
\def\Ds{\rlap{\, /}\;D }

\def\gym{g_{YM}}

\def\adot{\dot{a}}
\def\bdot{\dot{b}}
\def\bpa{\bar{\pa}}

\newcommand{\bes}{\begin{subequations}}
\newcommand{\ees}{\end{subequations}}

\newcommand{\eda}{\bar{\eta}}
\newcommand{\teta}{\tilde{\eta}}
\newcommand{\tta}{\tilde{\eta}}
\newcommand{\sP}{/\!\!\!\partial}
\newcommand{\sF}{\: /\!\!\!\! F}

\newcommand{\dd}{{d}}

\newcommand{\dsfrac}{\displaystyle\frac}

\begin{document}
\overfullrule=0pt
\parskip=2pt
\parindent=12pt
\headheight=0in \headsep=0in \topmargin=0in
\oddsidemargin=0in

\vspace{ -3cm}
\thispagestyle{empty}



 \vspace{0.1cm}

\setcounter{equation}{0}
\setcounter{footnote}{0}
\setcounter{section}{0}





\begin{center}

{\Large\bf $N^2$ entropy of 4D $\N=4$ SYM
  }

\vskip 0.8cm

I. Y. Park
\\

{\it Center for Quantum Spacetime, Sogang University\\
Shinsu-dong 1, Mapo-gu, 121-742 South Korea \\
}

and

\vspace{0.1cm}
{\it Department of Natural and Physical Sciences,
Philander Smith College
                               \\
Little Rock, AR 72202, USA \\
inyongpark05@gmail.com
}

\end{center}

 \vspace{0.1cm}

 \begin{abstract}

\ni We employ localization technique to derive $N^2$ entropy scaling
of four dimensional $\N=4$ SYM theory.

\end{abstract}
\newpage

\setcounter{equation}{0}
\setcounter{footnote}{0}
\setcounter{section}{0}


\section{Introduction}

Localization technique \cite{Witten:1988ze} has been proven very powerful in current studies of supersymmetric gauge theories \cite{Pestun:2007rz}\cite{Kapustin:2009kz}. It has been used recently in \cite{Drukker:2010nc} and \cite{Hatefi:2012sy} to derive the $N^{3/2}$ and $N^3$ scalings of M2 and M5 branes respectively.
This technique can bring a convenient redistribution of degrees of freedom
over the various parts of the path integral when it comes to evaluating a partition function.
This redistribution occurs via the choice of the localizing lagrangian and susy parameters.

It has long been known that the near-extremal D3 brane supergravity solution
has a Bekenstein-Hawking entropy that scales as $N^2$ \cite{Klebanov:1996un}. Because of this, there was
wide anticipation that it should be possible to observe the same scaling behavior from the SYM side.
Although the $N^2$ behavior has been demonstrated using a free gas model \cite{Gubser:1996de}, the same has not been achieved in the context of 4D SYM with interactions taken into account. (See \cite{Kinney:2005ej} and \cite{Dolan:2007rq} for related discussions.)

 There were basically two reasons why the $N^2$ scaling could not be established so far in the full (i.e., interacting) $\N=4$ SYM theory context. Firstly, it takes a non-pertubative technique, and one such technique has only been developed recently \cite{Pestun:2007rz}. Secondly, the bosonic contribution and fermionic contribution would cancel because of susy, leading to a behavior that is subleading to $N^2$ (presumably a $\ln N$ type behavior).
The work of \cite{Pestun:2007rz} has provided a framework of localization in which these issues can be tackled. In this work, we show that there exists a particular localization procedure that induces a susy breaking effect, thereby, leading to the expected $N^2$ entropy behavior. Once the susy is broken, it is obvious that the $N^2$ scaling is rather generic (although the numerical part of the coefficient is regularization-dependent).

We consider $\N=4$ SYM in a four dimensional {\em flat} space and employ
localization technique to compute the ``entropy". (Our method should be applicable to other cases.)
The aforementioned redistribution of the degrees of freedom reduces the amount of supersymmetry in effect.
 More specifically, the manner in which localization technique is applied hereby introduces a supersymmetry breaking effect (we will have more on this later), and this is crucial for our derivation of $N^2$ scaling of entropy. The numerical coefficient depends on the regularization method as discussed below.

In the $\N=1$ description of $\N=4$ SYM (see, e.g.,\cite{Gates:1983nr}\cite{Weinberg3} for a review of
supersymmetric gauge theories), there are three $\N=1$ chiral multiplets, $(\f,\psi,\F),(\f',\psi',\F'),(\f'',\psi'',\F'')$, and one $\N=1$ vector multiplet, $(A_\m,\l,D)$.
One of the three chiral multiplets - which we take to be $(\f',\psi',\F')$ - belongs to the $\N=2$ gauge multiplet and the other two belong
to the $\N=2$ hypermultiplet.
All three chiral multiplets are in the adjoint representation, on an equal footing in terms of the $\N=1$ description.

Even though localization technique\footnote{
 One subtle issue in the enterprise of computing partitions functions is associated with Higgs vevs vs instanton moduli. When one considers the path integral
one must not integrate over the Higgs vevs since they are physical ``observables", whereas
one integrates over the instanton moduli. If one's goal is to compute the entropy, the Higgs vevs
should not be integrated over.} greatly simplifies the evaluation of partition functions in general,
an explicit evaluation typically requires non-trivial computations.
This is especially
true when the computation involves instanton contributions; after introducing localization
terms in the action and finding an extremum configuration, one expands the localizing action
around the extremum configuration. The resulting expression is quadratic
in the fluctuation fields but the computation is still not simple since the coefficients now involve the instanton configuration; the evaluation would require the use of index theorem.
We show below
that there is a localization procedure that requires minimal use of this theorem. (Moreover, the necessary use of index theorem is one that has already been known, as we will point out.) For this, we start with the observation that not all the formulations of $\N=4$ SYM are equally effective in evaluating the partition function: a formulation in which the susy transformations of the $\N=1$ chiral multiplets do not involve the fields in the $\N=1$ vector multiplet is much more effective.

Our strategy is as follows. We employ the $\N=4$ SYM formulation in which
each $\N=1$ chiral multiplet transforms within itself (the $\N=1$ vector multiplet also transforms within itself), as discussed
 in several textbooks. To be specific, we follow
 the notations and conventions of \cite{Weinberg3}.
We take one of the chiral multiplets in the $\N=2$ hyper multiplet for an illustration of the localization of the chiral
 multiplets. As a result of the localization of the three chiral multiplets, the $\N=2$ gauge part of the lagrangian can be evaluated independently of the chiral multiplets.
 Combined with the results in the literature, the localization leads to
 full evaluation of the partition function.

The rest of the paper is organized as follows. In section 2, we start with the
$\N=1$ action that becomes the ``on-shell" $\N=4$ action once the auxiliary fields are integrated out. Localization procedure is carried out with a localization action chosen as \rf{clocal}. We note that the $N^2$-scaling arises due to the mismatch between the bosonic and fermionic degrees of freedom of the localization action. This mismatch should originate from the symmetry-breaking effect of the localization action chosen. In section 3, we note by adding an extra localization term (given in \rf{flt}) that there exists a certain range of the symmetry breaking effects associated with the localization actions.
In section 4, we conclude with speculation on the degrees of freedom that are
responsible for the $N^2$ growth of the entropy.

\section{Localization}

In one commonly used formulation of $\N=4$ SYM, the supersymmetry transformation of the chiral multiplet fermions
involve gauge field, and this feature makes the evaluation of partition complicated.
This is because the fluctuation fields couple to the instanton background, and
one must perform the instanton sum at the final stage.
Pleasantly enough, this
complication can largely be avoided by using the $\N=4$ SYM formulation that was discussed, e.g., in \cite{Weinberg3}.
The key point is that the susy transformations of the $\N=1$ chiral multiplets act within themselves
in that formulation, and in particular do not involve the gauge fields in the $\N=1$ vector multiplet.

Below we will show that the localization of the $\N=2$
hyper supermultiplet leads to decoupling between the $\N=2$ vector multiplet and hyper multiplet.
This implies that the vector multiplet part of the
partition function can be evaluated separately, and
one can again rely on localization technique for this
evaluation. The evaluation
consists of two parts: "classical" part and quadratic part around the instanton configurations.
The quadratic part was already evaluated through index theorem. See, e.g., \cite{Dorey:2002ik}, \cite{Bianchi:2007ft} and \cite{Vandoren:2008xg} for reviews.
In particular, the setup of \cite{Vandoren:2008xg} can be viewed as part of the aforementioned localization
 procedure as we will discuss shortly.

Let us consider the following $\N=1$ action
\bea
\cL &=& -\fr14 f_{A\m\n}f_A^{\m\n}-\fr12 \overline{\l_A} (\Ds \l)_A +\fr12 D_AD_A\nn\\
   && -(D_\m \f)_A^* (D^\m \f)_A-(D_\m \f')_A^* (D^\m \f')_A -(D_\m \f'')_A^* (D^\m \f'')_A \nn\\
   && -\fr12 \overline{\psi_A} (\Ds \psi)_A  -\fr12 \overline{\psi'_A} (\Ds \psi')_A -\fr12 \overline{\psi''_A} (\Ds \psi'')_A \nn\\
   &&-C_{ABC}\f_B^*\f_CD_A -C_{ABC}\f_B'^*\f_C'D_A-C_{ABC}\f_B''^*\f_C''D_A \nn\\
   &&-2\sqrt{2}\, \mbox{Re}\;C_{ABC}\f_A ({\psi'}_{BL}^T \e\; {\psi''}_{CL})
    -2\sqrt{2}\, \mbox{Re}\;C_{ABC}({\l}_{AL}^T \e\; {\psi}_{CL}) \f_B^* \nn\\
   && -2\sqrt{2}\, \mbox{Re}\;C_{ABC}\f'_B ({\psi''}_{CL}^T \e\; {\psi}_{AL})
    -2\sqrt{2}\, \mbox{Re}\;C_{ABC}\f''_C({\psi'}_{BL}^T \e\; {\psi}_{AL})  \nn\\
   && +2\sqrt{2}\, \mbox{Re}\;C_{ABC}({\psi'}_{BL}^T \e\; {\l}_{AL}){\f'}_C^*
    +2\sqrt{2}\, \mbox{Re}\;C_{ABC}({\psi''}_{BL}^T \e\; {\l}_{AL}) {\f''}_C^*  \nn\\
  &&+2\sqrt{2}\,\mbox{Re}\; C_{ABC}\f_A\f_B'\F_C''
    +2\sqrt{2}\,\mbox{Re}\; C_{ABC}\f_A\f_B''\F_C'
    +2\sqrt{2}\,\mbox{Re}\; C_{ABC}\f_A'\f_B''\F_C\nn\\
  && +\F_A^*\F_A +\F_A'^*\F_A' +\F_A''^*\F_A''
    \la{n4}
\eea
where $(A,B,C)$ are the adjoint indices and $C_{ABC}$ is the structure constant. The $(L,R)$ indices denote, for example,
\bea
\psi_L=\fr{1+\g_5}{2}\psi,\quad   \psi_R=\fr{1-\g_5}{2}\psi
\eea
The action \rf{n4} - which has $\N=1$ susy as it stands - becomes the $\N=4$ action once the field equations
of the auxiliary fields $\F'$'s, $D$ are
substituted. We proceed with \rf{n4} for now, and come back
to the $\N=4$ aspect later.

The two $\N=1$ chiral multiplets in the $\N=2$ hypermultiplet are on an equal footing in terms of the $\N=1$ description (this is also true for the $\N=1$ chiral multiplet in the $\N=2$ vector multiplet), and we illustrate the procedure with $(\f,\psi,\F)$. We consider the following localization action for $\f,\psi,\F$,\footnote{One may also add an addition localization term
\bea
Q\Big(c\,\overline{\psi_{AR}}\;\;[ \sqrt{2}\,\pa_\n \f_A^* \g^\n \e_R ]\Big)+h.c.
\la{flt}
\eea
where $c$ is an arbitrary constant. The value of the coefficient $c$ should be associated with certain phases
of the theory.
We will focus on \rf{clocal} for now with further analysis of \rf{flt} later.
}
\bea
Q\Big(\overline{\psi_{AL}}\;\;[ \sqrt{2}\,\pa_\n \f_A \g^\n \e_L ]\Big)
+h.c. \la{clocal}
\eea
$Q$ is a susy generator that includes a susy parameter, $\e$.
One of the necessary steps is finding the extremum configuration; we choose a trivial configuration in which
 all the fields vanish.
The localization of the $\N=2$ vector multiplet part decouples now. For the localization of the $\N=2$ vector multiplet, let us choose the localizing term such that it reproduces the entire $\N=2$
action when acted on by supersymmetry
 transformation.
The existence of such a localizing term
 is guaranteed by the fact that the lagrangian forms a supermultiplet.
The contribution of the vector multiplet gives a trivial contribution to the path-integral as one can see as follows. (This is particularly easy to see in the symmetric phase where all the scalar vevs are set
to zero. It should also be true in the broken symmetry phases.)
Consider a localization action for the $\N=2$ vector part of \rf{n4}, and name it as ${\cal V}_{\N=2\; vector}$. We choose ${\cal V}_{\N=2\; vector}$ such that $Q{\cal V}_{\N=2\; vector}$ yields the action that results from \rf{n4} by setting all the hypermultiplet fields in \rf{n4} to zero:
$(\f,\psi,\F)=0, (\f'',\psi'',\F'')=0$. In other words $Q{\cal V}_{\N=2\; vector}$ is the pure $\N=2$ vector multiplet action.  As usual, one keeps only the quadratic part for the fluctuation part. One can use the result of \cite{Vandoren:2008xg} for the contribution of the fluctuation part;
it was reviewed in \cite{Vandoren:2008xg} that the one-loop partition function of the system is trivial.
Therefore, the remaining evaluation is the ``classical" part summed over instanton contributions.
This part was considered in \cite{Okuda:2010ke}, and shown to yield a trivial result.
One subtlety is that the action \rf{n4} has only $\N=1$ supersymmetry due to the presence
of the auxiliary fields. We will address this
issue shortly.

Let us now continue with the localization of the $\N=2$ adjoint hypermultiplet.
The first localization term \rf{clocal} can be written
\bea
Q\Big(\overline{\psi_{AL}}\;\;[ \sqrt{2}\,\pa_\n \f_A \g^\n \e_L ]\Big)=\overline{(Q\psi_{AL})}\; \sqrt{2}\,\pa_\n \f_A \g^\n \e_L+
\overline{\psi_{AL}}\;\sqrt{2}\,\pa_\n (Q\f_A )\g^\n \e_L
 \la{form2}
\eea
 The supersymmetry transformations of the chiral multiplet are
\bea
\d \psi_{AL} &=& \sqrt{2}\; \pa_\m {\f_A} \g^\m \e_R +\sqrt{2}\F_A \e_L \nn\\
  \d \f_A &=& \sqrt{2}\;\overline{\a_R} \, \psi_{AL}
\eea
Note that the term in the square bracket in \rf{form2} is the same as $\d \psi_{AL}$ except for $\e_R\ra \e_L$.
This choice brings certain redistribution of the degrees of freedom mentioned in the introduction, and is a crucial step in our localization procedure.

It is straightforward to show that the first term of \rf{form2} yields
\bea
 \overline{(Q\psi_{AL})}\;\; \sqrt{2}\,\pa_\n \f_A \g^\n \e_L
 = -2\,\pa_\m \f_A \pa^\m \f_A^*  \;\Big(\bar{\e} \fr{1+\g^5}{2} \e\Big)
 +\F_A^* \pa_\n \f_A \Big(\bar{\e}\g_\n {\g_5}\e\Big)
 \la{bl}
\eea
which, in turn, implies
\bea
&& \overline{(Q\psi_{AL})}\;\; \sqrt{2}\,\pa_\n \f_A \g^\n \e_L+h.c.\nn\\
=&& -2\,
 \pa_\m \f_A \pa^\m \f_A^*  \;(\bar{\e}  \e)
 +\F_A^* \pa_\n \f_A (\bar{\e}\g_\n {\g_5}\e)
   +\F_A\pa_\n \f_A^* (\bar{\e}\g_\n {\g_5}\e)^\dagger
 \la{bl}
\eea
The fermionic part of \rf{form2}, $\overline{\psi_{AL}} \sqrt{2}\,\pa_\n (Q\f_A )\g^\n \e_L$, can be re-expressed by using the following Fierz identity for
arbitrary Mayorana-Weyl spinors $s_1,s_2$:
\bea
s_{1\a}\bar{s}_{2\b} &=& -\fr{1}{4}\mathbf{1}_{\a\b}(\sb_2 s_1)-\fr14{{\g^\m}_{\a\b}}(\sb_2 \g_\m s_1)
    +\fr1{32}[\g_\m,\g_n]_{\a\b}(\sb_2 [\g^\m,\g^\n] s_1) \nn\\
    && +\fr14 (\g_5\g_\m)_{\a\b}(\sb_2 \g_5\g^\m s_1)-\fr14 (\g_5)_{\a\b}(\sb_2 \g_5 s_1)
\eea
where $(\a,\b)$ are the spinor indices. Two terms among the five resulting terms trivially vanish
due to $\fr{1-\g_5}{2}\fr{1+\g_5}{2}=0$, and another term vanishes due to
\bea
\overline{\e_R}\,\fr{1+\g_5}{2}\,[\g_\r,\g_\s]\e_L=0,
\eea
an identity that can be proven by using the grassmannian nature of the spinors and (anti-)symmetry properties of the gamma matrices.
Combining the results so far, the fermionic part of \rf{form2} leads to
\bea
\overline{\psi_{AL}}\;\; \sqrt{2}\,\pa_\n (Q\f_A )\g^\n \e_L &=&
 - \fr12 \overline{\psi_A}\; (1-\g^5)  \pas \psi_A\; (\bar{\e} \fr{1+\g^5}{2}\e)
\eea
which, in turn, implies
\bea
\!\!\!\!\!\!\!\!\overline{\psi_{AL}}\;\sqrt{2}\,\pa_\n (Q\f_A )\g^\n \e_L +h.c.
= - \fr12(\,\overline{\psi_A}\; [1-\g^5]  \pas \psi_A)\; (\bar{\e}\e)
\eea
To derive this, we have used
\bea
(\bar{\e}\e)^\dagger=(\bar{\e}\e)\quad,\quad (\bar{\e}\g_5\e)^\dagger=-(\bar{\e}\g_5\e)
\eea
We have chosen zero vevs for all the fields as mentioned previously.
Combining the results so far, one obtains the following localization action
for one of the chiral multiplets:
\bea
\cL= 2\,\f_A \pa^2 \f_A^* \;(\bar{\e}  \e)
-\f_A \pa_\n\F_A^*   \Big(\bar{\e}\g^\n {\g_5}\e\Big)
      - \f_A^* \pa_\n\F_A  \Big(\bar{\e}\g^\n {\g_5}\e\Big)^\dagger
      - \fr12(\,\overline{\psi_A}\; [1-\g^5]  \pas \psi_A)\; (\bar{\e}\e)\nn\\
      \la{Llocal}
\eea

Let us now compute the ``free energy" of the system \rf{Llocal} from which the ``entropy" can be determined.
We consider zero temperature as in \cite{Hatefi:2012sy}.\footnote{Works that employed the Feynman diagrammatic techniques of computing the free energy at finite temperature can be
found, e.g., in \cite{Fotopoulos:1998es,VazquezMozo:1999ic,Kim:1999sg}.}
The quantity that we are after is basically the exponent of the partition function.
The meaning of "free energy" is an analogue of the energy functional as discussed in \cite{peskin}, an analogue of the Helmholtz free energy of the Euclidean SYM.
The path-integral now consists of two parts: classical and quadratic \rf{Llocal}.
Since the extremum configuration is such that all the fields vanish, there is no distinction between the $\N=1$ action and the $\N=4$
action as far as the classical sector is concerned.
The presence of the auxiliary fields affects the computation of the quadratic sector\footnote{
Since the quadratic part was obtained only by using the $\N=1$ susy transformation rules,
we expect that it would take the same form in the complete off-shell $\N=4$ formulation.
} in a minor
way, as we will see below.
 Let us consider the analogous issue in the vector
multiplet part before we get to the explicit computation of the chiral sector. The action used in \cite{Vandoren:2008xg} was without the vector multiplet auxiliary field $D$.
Suppose one uses the action with the auxiliary field. The path-integral would again consist of the classical part and the
quadratic part. As mentioned above, the localization action is identical to the original action
(one keeps only the quadratic part later, of course). It is rather obvious
from the structure of \rf{n4} and the decoupling of the chiral sector that the presence of the auxiliary field only trivially affects the evaluation of the path-integral.

The result so far can be summarized as follows:
The extremizing configurations for the chiral multiplets in the hyper multiplet are such that $<\phi>=\f_0,\;\; \psi=0$.
We consider the symmetric phase without scalar vevs, $\f_0=0$; in this sector, the
$\N=2$ gauge multiplet part of the action decouples from the hyper multiplet part of the action.
The vector multiplet part yields a trivial result, and the remaining task
is to evaluate \rf{Llocal}.

Compared with the standard form of the fermionic kinetic term, the projection
operator $\fr{1-\g^5}{2}$ is present in \rf{Llocal}. This indicates that half of the fermionic degrees of freedom are removed by the projection operator. Therefore a mismatch between the bosonic and fermionic degrees of freedom has arisen, and this disparity should be responsible for the $N^2$ entropy. (Otherwise one would get a slower scaling such as $\ln N$ \cite{Kinney:2005ej}.)
The $N^2$-scaling trivially follows from
the fact that these fields are in the adjoint representation: when the two adjoint indices are contracted they
yield $N^2-1$ factor to which we now turn.

 To compare with the case featuring the standard fermionic term, let us first compute the partition function of
 the standard free system,
\bea
 \cL
=   -\pa^\m \f_A\,  \pa_\m \f_A^*
    -\fr12(\bar{\psi_A} \pas \psi_A)
    \la{la}
\eea
The partition function should be regulated. We put the system in a 4D box and choose a regularization in which one of the spatial directions is made periodic with periodicity $L$. The 4D volume is taken to be $V_4=L^4$. A discrete sum is carried out for the selected direction and dimensional regularization
is employed for the remaining three dimensions after going to a continuum limit.

Two cautionary remarks are in order. Firstly, although this is reminiscent of finite temperature field theory, we are considering zero temperature as stated in the previous footnote. In particular, we impose periodic boundary conditions for the fermionic fields as well. The selected direction need not be the time: one can choose any of the three spatial directions to impose the periodic boundary conditions (,and indeed we have chosen one of the spatial direction as stated above). We take this regularization for a heuristic purpose; one would have to take other regularizations to be more rigorous. For example, one may consider putting the system on a sphere. (As well-known, however, SYM on a curved manifold has conformal anomaly in general. This issue has not been properly addressed in the works of related literature that employ a sphere background.) One would still find the $N^2$ behavior, although the overall numerical coefficient would be different.
Secondly, the reason for selecting out one of the dimensions (time or any other dimension) is as follows. It would be more desirable to use dimensional regularization from the beginning (i.e., without any $L$-regulator) if one were dealing with a massive theory. (As a matter of fact, dimensional regularization was used in \cite{Hatefi:2012sy} in which effective mass terms were present.) Dimensional regularization is awkward when one deals with a vacuum bubble diagram (such as the partition function we are considering in this work) in a massless theory. It is because the following integral is taken to vanish:
\bea
\int d^4p \fr{1}{(p^2)^\w}=0
\eea
 where $\w$ is any number. Therefore, one-loop energy will vanish for any theory that has a propagator of the form
 $\fr{1}{k^2}$. For a supersymmetric theory, this implies that one-loop energy vanishes because the bosonic and fermionic contributions separately vanish. This is an undesirable feature since it obscures the cancelation among the bosonic and fermionic contributions that other regularization schemes would display. For this reason, we discretize the momentum associated with one of the selected spatial dimension, and use dimensional regularization for the remaining dimensions.

More specifically, consider adding a localizing term $QS_L$
\bea
 \int D\Phi\; \exp\Big({\fr{i}{\gym^2}S+i\fr{t}{\gym^2}QS_L} \Big)
     \label{taction}
=\exp\Big({\fr{i}{\gym^2}}{V_4\cal A}\Big)
\eea
where $V_4$ denotes the volume of the 4D box. $S$ is the SYM action and $S_L$ is the total localization action,
\bea
\overline{\psi_{AR}}\;\;[ \sqrt{2}\,\pa_\n \f_A^* \g^\n \e_R ]+\cdots
\eea
where $\cdots$ denotes the vector part.
The first term can be found in the left-hand side of \rf{form2}, and we will continue to be implicit about the vector part since it has been already taken care of.
$\Phi$ is a collective symbol for the fields; $t$ is a localization parameter.
The quantity that we have been referring to as the ``free energy" is ${\cal A}$, the analogue of the Helmholtz free energy.
The ``entropy" per unit volume, $s$, is then defined by\footnote{This may be taken as a part of our proposal on how to compute the entropy at zero temperature.}
\bea
s\equiv \fr{\pa {\cal A}}{\pa (g_{YM}^2)}
\eea
The path integral of \rf{la} yields the following partition function\footnote{The time direction has been Euclideanized for the bosonic part. The fermionic part is to be Euclideanized below.}
\bea
{\cal A} &=& {-(N^2-1)\,2\fr{V_4}{ L}\sum_{n=-\infty}^\infty \int \fr{d^3 \vec{p}}{(2\pi)^3}
 \ln\Big[\fr{(2\pi n)^2}{L^2}+\vec{p}^2  \Big] }  \nn\\
 &&+{(N^2-1)\,\fr22\fr{ V_4}{L}\sum_{n=-\infty}^\infty \int \fr{d^3 \vec{p}}{(2\pi)^3} \ln \Big(\g^0[i\g^0 (2\pi n/L)+i\g^i p_i ]\Big) }
\eea
where irrelevant factors have been suppressed and the factor $V_4/L$ has appeared
while taking the trace (see, e.g., ch. 16 of \cite{Weinberg3}). (The factor $1/L$ appeared while converting the integral over the selected spatial to the sum.) The factor 2 comes from the fact that there are three chiral multiplets and the factor $1/2$ in the fermionic part is due to the fact that the fermion is of Mayorana type.
Let us evaluate the exponents; the bosonic one is
\bea
h_B\equiv 2 \sum_{n=-\infty}^\infty \int \fr{d^3 \vec{p}}{(2\pi)^3}
 \ln\Big[\fr{(2\pi n)^2}{L^2}+\vec{p}^2  \Big]
\eea
Consider
\bea
\fr{\pa}{\pa (1/L)}h_B &=& \fr{32 \pi^2}{L} \sum_{n=1}^\infty  n^2 \int \fr{d^3 \vec{p}}{(2\pi)^3}
 \fr{1}{\fr{(2\pi n)^2}{L^2}+\vec{p}^2  }
\eea
Let us turn to the fermionic exponent; define
\bea
 h_F &\equiv & 2\cdot \fr12 \sum_{n=-\infty}^\infty \int \fr{d^3 \vec{p}}{(2\pi)^3} \ln \Big(\g^0[i\g^0 (2\pi n /L)+i\g^i k_i ]\Big) \nn\\
 &=&  \sum_{n=-\infty}^\infty \int \fr{d^3 \vec{p}}{(2\pi)^3} \ln \Big(-i (2\pi n /L)+i\g^0\g^i k_i ]\Big)
\eea
Taking the $1/L$-derivative yields
\bea
 \fr{\pa h_F}{\pa (1/L)} &\equiv &  \Tr \sum_{n=-\infty}^\infty(-2\pi i n) \int \fr{d^3 \vec{p}}{(2\pi)^3} \fr{1}{ \Big(\g^0[i\g^0 (2\pi n/ L)+i\g^i k_i ]\Big)} \nn\\
 &\Rightarrow & 32\pi^2 L   \sum_{n=1}^\infty  n^2\int \fr{d^3 \vec{p}}{(2\pi)^3} \fr{1}{(2\pi n /L)^2+\vec{k}^2}
 \la{be}
\eea
where `$\Rightarrow$' indicates the Euclideanization mentioned previously.
This cancels the bosonic contribution exactly as it should due to the supersymmetry.

Let us turn to the evaluation of \rf{Llocal}. It is essentially the difference in the structures of \rf{Llocal} and \rf{la} that leads to non-vanishing entropy in the case of \rf{Llocal}.
Let us consider the path-integral over $\f$ by using
\bea
\int \fr{dz d\zb}{2\p i}e^{-\zb A z+\bar{u}z+u\zb}=
(\det A)^{-1}e^{\bar{u}A^{-1}u}
\eea
The factor that contains the auxiliary field $\F_A$ is
\bea
e^{-(\pa_\m\F_A)\fr{1}{2\pa^2}( \pa_\n\F_A^*) \;   (\bar{\e}\g^\m {\g_5}\e) (\bar{\e}\g^\n {\g_5}\e)}
\eea
Due to the following identity \cite{Weinberg3}
\bea
(\bar{\e}\g^\m {\g_5}\e) (\bar{\e}\g^\n {\g_5}\e)=-\eta^{\m\n}(\bar{\e}{\g_5}\e)^2
\eea
the path-integral over $\F$ only produces an irrelevant (i.e., momentum independent) factor. Therefore, one may effectively consider
\bea
\cL= 2\,\f_A \pa^2 \f_A^* \;(\bar{\e}  \e)
      - \fr12(\,\overline{\psi_A}\; [1-\g^5]  \pas \psi_A)\; (\bar{\e}\e)
      \la{Llocal2}
\eea
instead of \rf{Llocal}.
One gets similar expressions for the bosonic and fermionic contributions such as those that belong to \rf{la} (with differences only in the overall numerical coefficients), and can explicitly evaluate the expressions that have appeared above. For example,
\bea
 \fr{\pa}{\pa (1/L)}h_B &=& \fr{32\pi^3}{ L^2} \G\Big(-\fr12\Big)\sum n^3 = -16 \pi^{2}L^2 \z(-3)=-\fr{2\pi^2}{15} \fr1{L^2} \la{phb}
\eea
where we have used a formal identity, $\G(-1/2)=-2\G(1/2)$, which should be viewed as part of the regularization.
The equality follows from the combined use of dimensional regularization (or regularization by dimensional reduction) and zeta function regularization.\footnote{$\z$-function regularization requires more care to be compatible with gauge invariance \cite{Rebhan:1988ed}. We put this issue aside, since, for any reasonable regularization, the difference in the structures of \rf{la} and \rf{Llocal2} should be responsible for the non-vanishing entropy of \rf{Llocal2}, as stated above.
Other regularizations should be possible; for example, one
may put the system on a sphere. (See, e.g., \cite{Russo:2012ay}.) One would still find the $N^2$ behavior although the overall numerical coefficient would be different. As well-known, however, SYM on a curved manifold has conformal anomaly in general. As far as we can tell, this issue has not been properly addressed in the works that employ a sphere background.
}
``$\z$" denotes the Riemann zeta function, and we have used $\z(-3)=\fr{1}{120}$.
It follows from \rf{phb}
\bea
h_B=-\fr{2\pi^2}{45}\fr1{ L^3}
\eea
Similarly, one can show
\bea
h_F=\fr{\pi^2}{45} \fr1{L^3}
\eea
From these results, one finds
\bea
{\cal A}= (N^2-1)g_{YM}^2\;\fr{\pi^2}{45}
\eea
which, in turn, implies that the entropy per unit volume is
\bea
s=\fr{\pi^2}{45}  (N^2-1)
\eea
It would be interesting to repeat the computation by adopting a different regularization, and obtain
the numerical coefficient that one would get in place of $\fr{\pi^2}{45}$.

\section{Localization actions and symmetry breaking}

Let us explore the symmetry breaking patterns further by adding \rf{flt}:
\bea
Q\Big(\overline{\psi_{AL}}\;[ \sqrt{2}\,\pa_\n \f_A \g^\n \e_L ]
+c\;\overline{\psi_{AR}}\;[ \sqrt{2}\,\pa_\n \f_A^* \g^\n \e_R ]\Big)
+h.c. \la{clocal2}
\eea
where $c$ is an arbitrary constant that will be associated with the breaking patterns.
Following the steps analogous to the ones in the previous section, one can show that the bosonic part of the second
term in \rf{clocal} yields
\bea
&&c\,Q\Big(\overline{\psi_{AR}}\;[ \sqrt{2}\,\pa_\n \f_A^* \g^\n \e_R ]\Big)
 +h.c. \nn\\
=&& -2c\,
 \pa_\m \f_A \pa^\m \f_A^*  \;(\bar{\e}  \e)
- \fr{c}2(\,\overline{\psi_A}\; [1+\g^5]  \pas \psi_A)\; (\bar{\e}\e)
   \la{Lextra}
\eea
where we have omitted the irrelevant $\F$-terms.
Now one can carry out the path-integral; obviously the coefficient
of $N^2$ would contain the coefficient $c$. Although the precise physical meaning of the coefficient $c$ will require more work, we believe that
it is likely to be a non-perturbative analogue of the scalar vevs.\footnote{
The special case of $c=1$ leads to vanishing entropy. In this sense, $c=1$ may be associated with a more symmetric phase of the theory. Possibly, a maximization procedure over $c$ may be necessary.
(The issue of the cohomological vacuum choice and the corresponding gauge-fixing should be examined
to better understand the origin of the $c$-dependence.)
}

\section{Conclusion}

On account of cancellation between the bosonic and fermionic degrees of freedom,
it is puzzling to some extent that $\N=4$ SYM - a theory with a large amount of supersymmetry - has led to
$N^2$ entropy. Furthermore, the subleading terms in $N$ do not exist in our result. We believe that the arrival of the $N^2$ entropy should be attributed to a
supersymmetry breaking effect that the localization technique has brought through the redistribution of degrees of freedom that we discussed on in the introduction. Recall that, in localization technique, it is the supersymmetry of the localization lagrangian that is relevant for the purpose, rather than that of the original lagrangian.
Obviously, the localization lagrangian has much less supersymmetry than the original lagrangian.
Genuine finite temperature effects and/or the dielectric effect (as discussed in \cite{Hatefi:2012sy}) are likely to induce subleading contributions in $N$.

Finally we comment on the degrees of freedom that should be responsible for the entropy.
It is likely to be the goldstino multiplet of the SYM counterparts of the supergravity $\fr1{16}$-BPS states that
are behind the $N^2$ behavior.
Some discussions on $\fr1{16}$-BPS states can be found in \cite{Biswas:2006tj} and \cite{Kim:2006he}. We leave the better understanding of this issue and the issue of symmetry breaking patterns discussed in section 3 for future endeavors.

\vspace{.8in}

\ni {\bf Acknowledgements}\\
I am grateful to M. Rocek for the valuable discussions. I thank
E. Hatefi for his hospitality during my visit to ICTP, Trieste.

\newpage

\renewcommand{\theequation}{A.\arabic{equation}}
\setcounter{equation}{0}
\section*{Appendix A: conventions and useful formulas }

We summarize the conventions and some of the useful identities.
The susy transformations of the chiral multplet are given by
\bea
\d \psi_L &=& \sqrt{2}\; \pa_\m \f \g^\m \e_R +\sqrt{2} {\cal F} \e_L \nn\\
  \d{\cal F} &=& \sqrt{2}\;\overline{\e_L} \pas \psi_L   \nn\\
  \d \f &=& \sqrt{2}\;\overline{\e_R}  \psi_L
\eea
and
\bea
\d \psi_R &=& \sqrt{2}\; \pa_\m \phi^* \g^\m \e_L +\sqrt{2} {\cal F}^* \e_R \nn\\
  \d{\cal F}^* &=& \sqrt{2}\;\overline{\e_R} \pas \psi_R   \nn\\
  \d \f^* &=& \sqrt{2}\;\overline{\e_L}  \psi_R
\eea
The `bar' on a spinor such as $\sb$ is defined by
\bea
\bar{s}\eqv s^\dagger \b =s^T \e \g_5
\eea
with
\bea
\e_{\a\b}=\left(
\begin{array}{cc}
e & 0 \\
0 & e \\
\end{array}
\right),\quad
e=\left(
\begin{array}{cc}
0 & 1 \\
-1 & 0 \\
\end{array}
\right)
\eea
where $\a,\b$ are 4D spinor indices. The matrix $\e$ satisfies
\bea
 [\e,\g_5]=0, \quad \e^2=-1
\eea
The following identities were used frequently in the main body,
\bea
M^T =\left\{
\begin{array}{c}
{\cal C}M {\cal C}^{-1}\quad M=1,\g_5,\g_5\g_\m \\
-{\cal C}M {\cal C}^{-1}  \quad M=\g_\m, \g_{\m\n}
\end{array}
\right.
\eea
where ${\cal C}$ is the charge conjugation matrix, and it satisfies
\bea
&& {\cal C} \eqv \g_2 \b=-\e \g_5,\quad
{\cal C}^{-1} =\e \g_5
\eea

\newpage

\end{document}